\documentclass[reprint,fleqn,superscriptaddress,twocolumn,showpacs,
amsmath,amssymb,prb,aps,bibliography]{revtex4-1}
\usepackage[all]{xy}
\usepackage{gensymb}
\usepackage{graphicx,psfrag,times,epsfig,color}
\usepackage{verbatim,natbib}

\usepackage{color}
\usepackage{makeidx}
\usepackage{amsmath}
\usepackage{bm}
\usepackage{amsfonts}
\usepackage{amssymb}
\usepackage{hyperref}

\begin{document}

\title{Why the superfluid density tracks 
{\it T}$_{\rm c}$ in cuprate superconductors?}
\author{E. V. L. de Mello}
\affiliation{Instituto de F\'{\i}sica, Universidade Federal Fluminense, 24210-346 Niter\'oi, RJ, Brazil}

\email[Corresponding author: ]{evandro@if.uff.br}

\begin{abstract}

One of the first finding concerning the superconducting (SC) density $n_{\rm sc}$ in 
cuprates was their small magnitudes that revealed the importance of phase fluctuations. 
More recently, measurements in a variety of overdoped cuprates 
indicate that it is also much smaller than expected from BCS theories and 
falls smoothly to zero as doping is increased.
We explain these observations by an electronic phase separation theory with
a Ginzburg-Landau                                                                                                                                                                                                                                                                                                                                                      potential $V_{\rm GL}$ that produces alternating charge domains whose fluctuations
lead to localized
SC order parameters that are
connected by Josephson coupling $E_{\rm J}$.  The average ${\left <E_{\rm J}( p,T)\right>}$
is proportional to the local superfluid phase stiffness 
$\rho_{\rm sc} \propto n_{\rm sc}$
.
The fraction of condensed carriers decreases in the overdoped region due to the 
weakening of $V_{\rm GL}$. The results agreed with $\rho_{\rm sc}(p)$ vs. 
$T_{\rm c}(p)$ and the Drude-like peak measurements.

\end{abstract}

\maketitle

Almost thirty years ago Uemura {\it et al}\cite{Uemura1989} performed a series of $\mu$-SR experiments 
and found a universal linear scaling between $T_{\rm c}(p)$ and the zero temperature superfluid 
density $n_{\rm sc}(p,0)$ at low $p$. The proportionality law was not confirmed by
subsequent experiments that measured a slightly parabolic behavior, but all measurements
agreed with the saturation of $n_{\rm sc}(p,0)$ at or beyond optimum 
doping\cite{LscoOver1993,Tallon2003,Chu2003,Randeria2011}.
Another general observation was the low superfluid densities
with one to two orders of magnitude less than of
conventional BCS superconductors\cite{Uemura1989,LscoOver1993,Tallon2003,Chu2003,Randeria2011}.
The small magnitudes of $n_{\rm sc}$ are
indicative that phase fluctuations play a significant role in the physics
of cuprates\cite{Emery1995,Bilbro2011}. One possible implication is that
Cooper pair formation occurs at some onset
temperature, but long-range phase coherence does not occur until the 
temperature is lowered to $T_{\rm c}$. In fact, many experiments have measured 
persisting SC correlations\cite{Gomes2007,Kanigel2008,Dubroka2011}
and diamagnetic responses\cite{Rigamonti2003,Ong2010,FVidal2013} at temperatures well above $T_{\rm c}$.

Recently  Bo$\breve{z}$ovi\'c {\it et al}\cite{Bozovic2016} 
found a similar linear scaling relation on overdoped La$_{\rm 2-x}$Sr$_{\rm x}$CuO$_4$
(LSCO) films but  with a negative slope, in close agreement with
earlier works\cite{Uemura1993,LscoOver1993} on Ti$_2$Ba$_2$CuO$_{6+\delta}$(Tl2201).
Putting all experiments together, the dominant picture is an almost 
linear increase of $n_{\rm sc}(p,0)$ in the 
underdoped region that saturates right after optimum doping and decreases with $T_c(p)$ going down
to zero in the far overdoped samples. This decreasing of $n_{\rm sc}(p,0)$
after saturation was not expected because  
doping brings more charges to the CuO planes what, in principle would increase the
number of Cooper pairs in a standard BCS framework of superconductivity. 
A subsequent experiment\cite{Bozovic2019} found that a significant 
fraction of the carriers remains
uncondensed in a wide Drude-like peak as $T \rightarrow 0$, while 
$n_{\rm sc}(p,0)$ remains proportional to $T_{\rm c}(p)$ and vanishes in the limit of superconductivity.
This experiment\cite{Bozovic2019}
shows that overdoped superconductors behave like a two fluids system and 
helped to understand the puzzle of the ``missing'' carriers.

After all these years there are not a wide accepted theory to the superfluid densities
in cuprates. Most likely because the ubiquitous presence of 
incommensurate charge ordering (CO)\cite{Comin2016} on cuprates
introduced a degree of complexity difficult to be incorporated in any realistic theory. 
On the other hand, overdoped materials have been believed to be well described
by Fermi liquid theory\cite{Keimer2015} and to have a SC state with
conventional BCS-like properties.
While a
model considering pair-breaking due to impurity scattering in a BCS like $d$-wave
superconductor reproduced well the $T_c(p)$ vs. $n_{\rm sc}(p,0)$\cite{Broun2017},
it is
inconsistent with the recent observation of a wide Drude peak\cite{Bozovic2019}.
A subsequent work\cite{Broun2018} reconciled the observed broad residual Drude peak
with the behavior of $n_{\rm sc}(p,0)$ within the Born limit. However,
an infinite number of weak scatters center (Born limit) is not reasonable
considering that the scattering rate of the $T \rightarrow 0$ residual Drude is 
about the same of the normal state\cite{Bozovic2019}.

In this letter we provide a unified explanation to the old results\cite{Uemura1989},
the suppression in superfluid densities in the overdoped experiments\cite{Bozovic2016}
and to the residual Drude contribution\cite{Bozovic2019}. The calculations 
follow the method of the preceding paper\cite{Mello2019a} and in earlier 
works\cite{DeMello2012, DeMello2014,Mello2017,Mello2019a} which starting point is the simulations
of charge instabilities like stripes,
incommensurate charge order (CO) or charge density waves 
(CDW)\cite{Wise2008,Wu2011,Chang2012,Blanco-Canosa2014,Huecker2014,DaSilvaNeto2014,Comin2014,Campi2015,
Comin2015a,Comin2016,Tabis2017}. It is important to emphasize that some 
of these experiments\cite{Chang2012,Comin2014} provide strong indications that the
pseudogap (PG)
is correlated to the charge instabilities and this was confirmed by
specific calculations\cite{Mello2019a}. Based on these results,
we apply the method to the overdoped regions where CO is more difficult to 
be detected.

In short, to define the important variables and parameters, our approach is based on
the time-dependent Cahn-Hilliard (CH) nonlinear differential equation via
a Ginzburg-Landau (GL) free energy expansion in terms of a diffusive or phase separation
order parameter $u(r_i,t)$ associated with the local hole density $p(r_i,t)$
that evolves in time $t$. As discussed above,
at $T \sim T^*$, the GL free energy potential $V_{\rm GL}(r_i,t)$ starts to segregate
the charge wave functions in superlattices formed by alternating hole-rich and hole-poor domains, 
with different CO wavelength $\lambda_{\rm CO}$ and structures. 
The non-uniform charge distribution interacts with the Cu atoms electronic clouds
and may induce local SC pairing interactions. 
This fundamental point is discussed in some detail in Ref.[\onlinecite{Mello2019a}].

To calculate $T_{\rm c}(p)$ we have performed Bogoliubov-deGennes calculations
on various CO density maps\cite{DeMello2014,Mello2019a,Mello2019a} what yields local
amplitudes $\Delta_d(r_i)$ with the same wavelength $\lambda_{\rm CO}$.
The localized $\Delta_d(r_i)$ is in agreement
with SC coherence lengths $\xi$ typically smaller\cite{ECarlson2002} 
than average $\lambda_{\rm CO}$\cite{Comin2016}, what implies that the charge
domains may behave as mesoscopic SC grains.
This gives rise to Josephson coupling between the distinct SC regions with
energy $E_{\rm J}(r_{ij})$ that is the lattice version of the 
local superfluid density $\rho_{\rm sc}$\cite{Spivak1991}. 
We have already explained\cite{Mello2017} that
for two $d$-wave superconductors junction is sufficient to use the following 
$s$-wave relation for the average Josephson coupling energy\cite{AB1963}:

\begin{equation}
 {\left < E_{\rm J}(p,T) \right >} = \frac{\pi \hbar {\left <\Delta_d(p,T)\right >}}
 {2 e^2 R_{\rm n}(p)} 
 {\rm tanh} \bigl [\frac{\left <\Delta_d(p,T)\right >}{2k_{\rm B}T} \bigr ] ,
\label{EJ} 
\end{equation}
where $R_{\rm n}(p)$ are taken to be proportional to the 
$T \gtrapprox T_{\rm c}$ normal-state in-plane resistivity 
$\rho_{ab}(p)$ obtained from typical $\rho_{ab}(p,T)\times T$ curves for Bi2212,
LSCO and Y123\cite{Ando2004}. The values for Bi2201 are the same used in 
Ref. [\onlinecite{Mello2017}]. The proportionality constant between $R_{\rm n}$
and $\rho_{ab}$ is found matching the optimal $T_{\rm c}(p=0.16)$.
The same constant is used between all
$\rho_{ab}(p)$ and $R_{\rm n}(p)$ and we list some $R_{\rm n}(p)$ in Table I.
The  spatial average
${\left < \Delta_d(p,T) \right >} \equiv \sum_i^N \Delta_d(r_i,p,T)/N$ contains
all the planar sites.


\begin{figure}[!ht]
 \includegraphics[height=6.0cm]{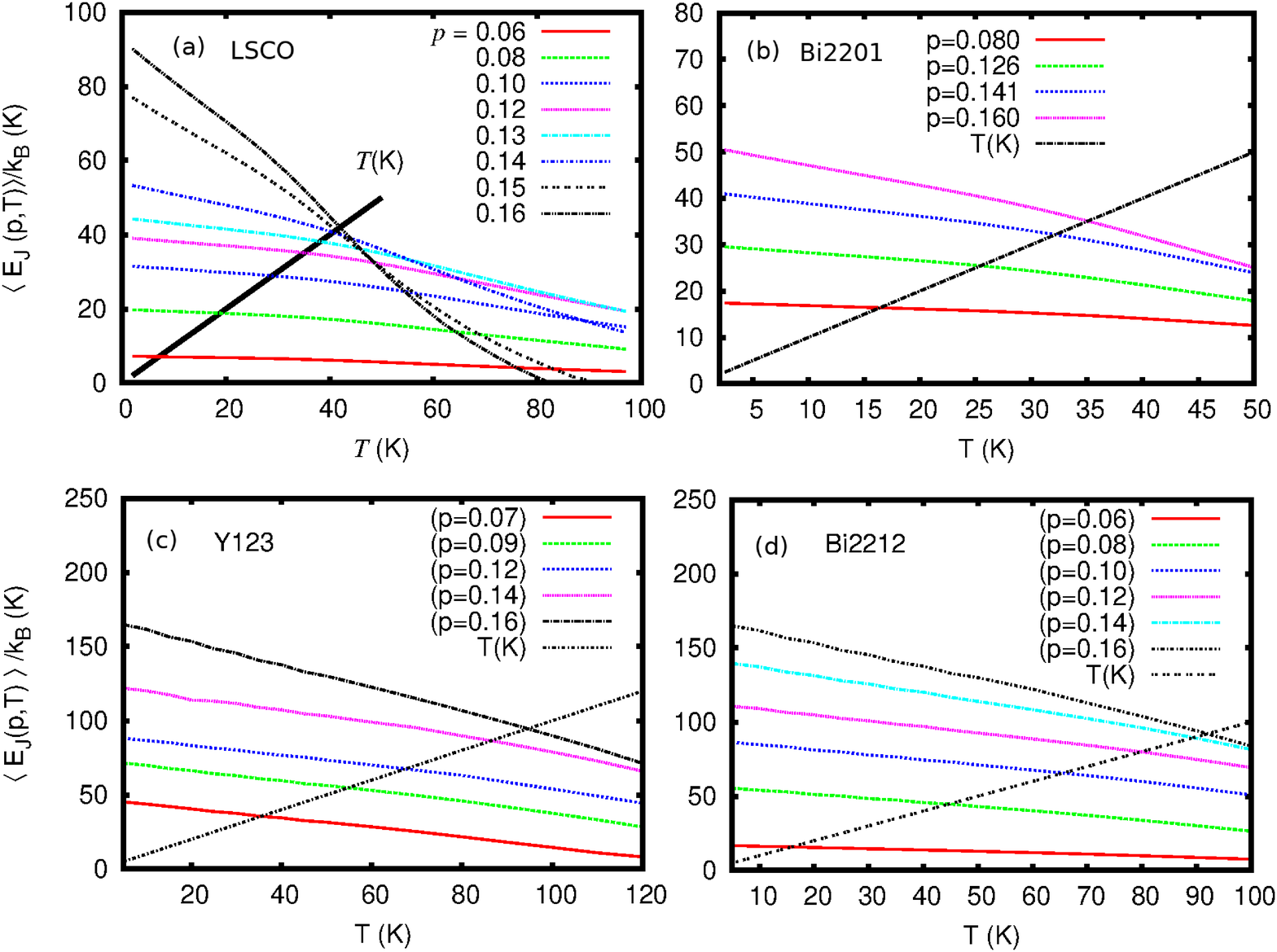}
\caption{The calculations of ${\left < E_{\rm J}(p,T) \right >}$ as function 
of temperature $T$ and the straight line $k_{\rm B}T$. Their intersection
determines the onset of long range order, i.e., $T_{\rm c}(p)$. 
The superfluid density $\rho_{\rm sc} (0)$ is given by 
${\left < E_{\rm J}(p,0) \right >}$. 
}
\label{figEJxT}
\end{figure}

\begin{figure}[!ht]
 \includegraphics[height=4.0cm]{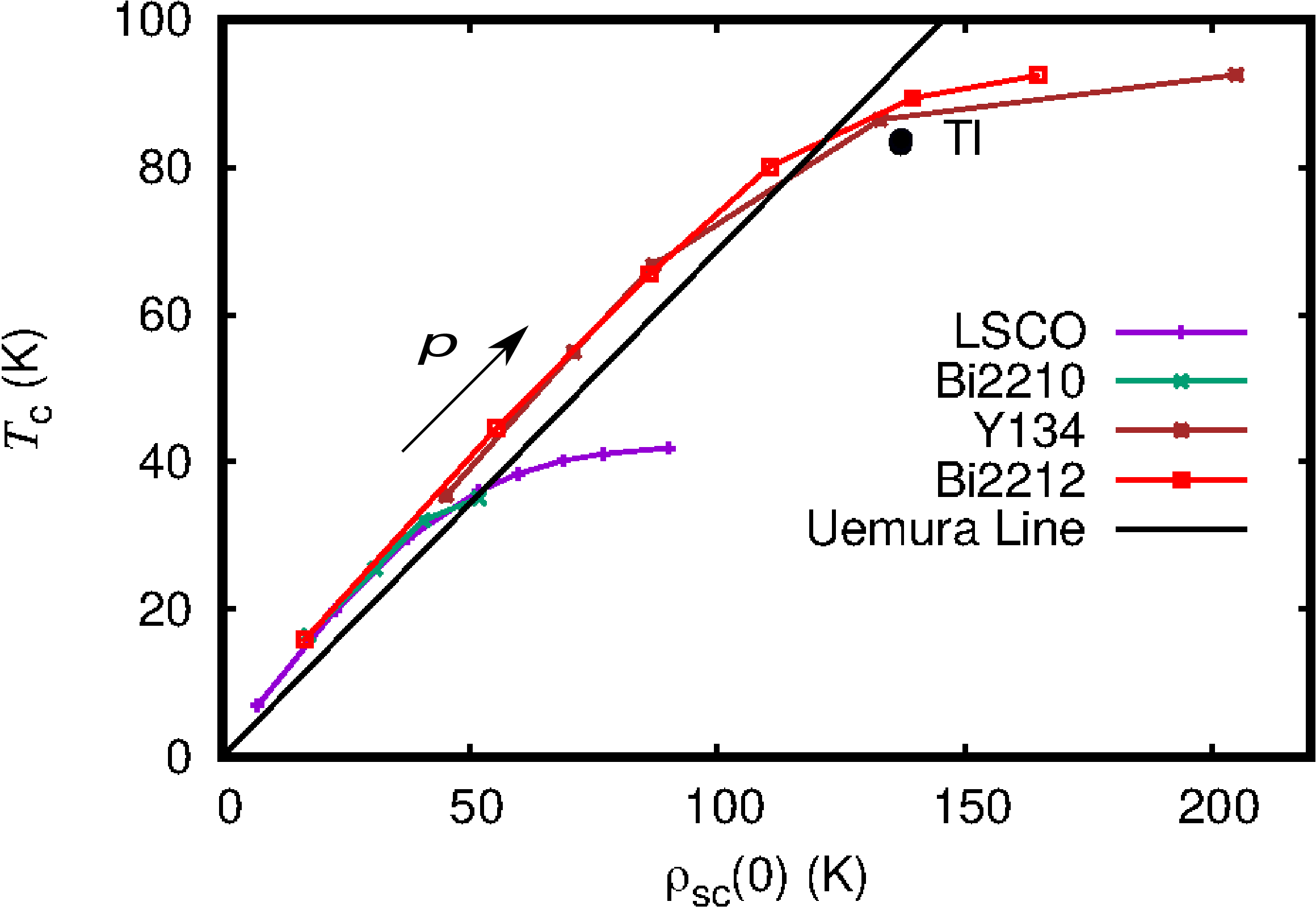}
\caption{ 
The values of  $T_{\rm c}(p)$ and $\rho_{\rm sc}(p,0)$
derived directly from the plots of Fig. \ref{figEJxT} and 
listed in Table I. The curves are not linear at low $p$ as measured
by Ref.[\onlinecite{Uemura1989}]  but close to 
experimental results of Ref.[\onlinecite{LscoOver1993}]. A Tl2201 
slightly overdoped sample with $T_{\rm c}=84$K is included. We also draw
the linear Uemura proposal\cite{Uemura1989} and the arrow points to the 
direction of increase $p$.
}
\label{figTcxRhoU}
\end{figure}

We evaluate 
${\left < E_{\rm J}(p,T) \right >}$ for the 
Bi$_{{\rm 2-y}}$Pb$_{\rm y}$Sr$_{\rm 2-z}$La$_{\rm z}$CuO$_{6+\delta}$ (Bi2201), LSCO,
Bi$_2$Sr$_2$CaCu$_2$O$_{8+\delta}$ (Bi2212) and 
YBa$_2$Cu$_3$O$_{6+\delta}$ (Y123) systems
and show the plots in Fig. \ref{figEJxT}(a-d). 
They contain the two important quantities to this work: The low temperature 
superfluid phase stiffness $\rho_{\rm sc} (p,0) (\equiv {\left < E_{\rm J}(p,0) \right >})$ 
that is read directly at $T=0$ K and
$T_{\rm c}(p)$ that is obtained from the intersection between the Josephson coupling
${\left < E_{\rm J}(p,T) \right >}$ and the thermal disorder energy $k_{\rm B}T$. The 
results for $T_{\rm c}(p)$ and $\rho_{\rm sc}(p,0)$ for four underdoped cuprates 
are listed in Table I and plotted in Fig. \ref{figTcxRhoU}. 
We draw also the Uemura line for reference that shows that our 
points are closer to a parabola than a straight line in better agreement with the
LSCO and Y123 results\cite{LscoOver1993,Tallon2003}. We also plot a slightly
overdoped Tl2201 with $T_{\rm c}(p)$\cite{LscoOver1993} that is close to the 
optimal points. 

The behavior of the $T_{\rm c}(p)$ vs. $\rho_{\rm sc}(p,0)$ may be understood by
examining the Josephson energy curves ${\left < E_{\rm J}(p,T) \right >}$ vs. $T$  
plotted in Figs. \ref{figEJxT}. In the underdoped region,
$\langle \Delta_{d}(p,T)\rangle$ is almost constant at low $T$ 
and vanishes much above $T_{\rm c}(p)$. 
This implies that ${\left < E_{\rm J}(p,T) \right >}$ vs. $T$ are slowly decreasing 
approximate straight lines (see Figs. \ref{figEJxT}(a-d))
establishing a direct relation between ${\left < E_{\rm J}(p,0) \right >} = \rho_{\rm sc} (p,0)$
and ${\left < E_{\rm J}(p,T_{\rm c}) \right >} = k_{\rm B}T_{\rm c}(p)$
that is not strictly linear but is close to Uemura's originated proposal
(see Fig. \ref{figTcxRhoU}).
When $p$ gets near the opitimum doping $p_{\rm opt}$, 
$T^*(p)$ and $\langle \Delta_{d}(p,T)\rangle$ diminish with $T$
and ${\left < E_{\rm J}(p,T_{\rm c}) \right >}$ vs. $T$ decrease faster
and some series saturates near $p \ge 0.13$.
On the other hand, the Bi2201 series have very steep $T^*(p)$ vs. $p$ and their 
${\left < E_{\rm J}(p,T_{\rm c})\right>}$ are almost horizontal lines and
does not saturate up the optimal doping.
Thus, whenever $T^*(p)$ becomes closer to $T_{\rm c}(p)$, 
${\left < E_{\rm J}(p,T_{\rm c}) \right >}$ will have steeper slopes and 
$\rho_{\rm sc}(p,0)$ vs. $T_{\rm c}(p)$ saturates and may decrease with $p$. This behavior 
was observed in the cuprates studied by Uemura {\it et al}\cite{Uemura1989}.

\begin{table}[!ht] 
\caption{Selected values of $T_{\rm c}(p)$ estimated from the Josephson
couplings given in Fig. \ref{figEJxT} for the four cuprate families. 
The second lines give the low temperature superfluid densities $\rho_{\rm sc} (0)$. 
The third lines give the normal resistivity just above $T_{\rm c}(p)$ that enters
in Eq. (\ref{EJ}) and are proportional to the experimental\cite{Ando2004}
measured values of $\rho_{ab}$ just above $T_{\rm c}$. $V_{\rm GL}(p = 0.16)$
(or $T_{\rm c}(0.16)$) and $B$ from 
$R_{\rm n} = B \times \rho_{\rm ab}(p = 0.16)$  
are the only two adjustable parameters of each series and are in bold blue. 
}
\centering
\begin{tabular}{|c|c|c|c|c|c|c|}\hline \hline

  $p $ (LSCO) & 0.06  & 0.08 & 0.10  & 0.12 & 0.14  &  0.16    \\ \hline
 $V_{\rm GL}$ (meV) &{\color{red} 318} & {\color{red}294 }& {\color{red}275} 
& {\color{red}267 } & {\color{red}248 }& {\color{blue}{\bf 234}} \\ \hline
$\langle \Delta_{d}(0K)\rangle$ & {\color{red}18.7} & {\color{red} 16.2} & {\color{red} 18.1 }
& {\color{red} 18.3} &{\color{red} 17.5} &{\color{red} 16.8} \\ 
  (meV)  & &  &  &  &  & \\ \hline
$T_{\rm c}$ (K)  & {\color{red} 6.7} & {\color{red} 19.5} & {\color{red} 29.4} 
&{\color{red} 35.4} &{\color{red} 39.8} & {\color{blue} {\bf 41.9}} \\ \hline
$\rho_{\rm sc}(0)$(K) & {\color{red} 7.0} & {\color{red} 25.0} & {\color{red} 38.5} 
& {\color{red} 44.5} & {\color{red} 60.6} & {\color{red} 90.1} \\ \hline
$R_{\rm n}$ (m$\Omega$cm)& {\color{red} 0.790} & {\color{red} 0.281} &{\color{red} 0.175}
&{\color{red} 0.127} &{\color{red} 0.092} & {\color{blue}{\bf 0.067}} \\
\hline \hline
  $p $ (Bi2201) & 0.114  & 0.126 & 0.141  & 0.16 &   &     \\ \hline
$T_{\rm c}$ (K)  & {\color{red} 16.5} & {\color{red} 25.5} & {\color{red} 32.2} 
&{\color{blue}{\bf 35.2}} &  &  \\ \hline
$\rho_{\rm sc}(0)$ (K) & {\color{red} 14.4} & {\color{red} 30.0} & {\color{red} 41.0} 
& {\color{red} 50.4} &  &  \\ \hline
$R_{\rm n}$ ($\mu\Omega$cm)& {\color{red} 52.5} & {\color{red} 31.7} &{\color{red} 24.2}
&{\color{blue}{\bf 18.3}} &  &  \\
\hline \hline
  $p $ (Y123) & 0.07  & 0.09 & 0.12  & 0.14 &  0.16 &     \\ \hline
$T_{\rm c}$ (K)  & {\color{red} 35.5} & {\color{red} 55.0} & {\color{red} 66.7} 
&{\color{red} 86.5} & {\color{blue}{\bf 92.6}} &  \\ \hline
$\rho_{\rm sc}(0)$ (K) & {\color{red} 45.2} & {\color{red} 71.0} & {\color{red} 87.1} 
& {\color{red} 121.9} & {\color{red} 166.0 } &  \\ \hline
$R_{\rm n}$ ($\mu\Omega$cm)& {\color{red} 150} & {\color{red} 80} &{\color{red} 50}
&{\color{red} 45} &{\color{blue}{\bf 40}} &  \\
\hline \hline
  $p $ (Bi2212) & 0.06  & 0.08 & 0.10  & 0.12 & 0.14  &  0.16    \\ \hline
$T_{\rm c}$ (K)  & {\color{red} 15.9} & {\color{red} 44.6} & {\color{red} 64.6} 
&{\color{red} 80.1} &{\color{red} 89.40} & {\color{blue}{\bf 92.5}} \\ \hline
$\rho_{\rm sc}(0)$ (K) & {\color{red} 16.7} & {\color{red} 55.35} & {\color{red} 89.8} 
& {\color{red} 110.6} & {\color{red} 139.5} & {\color{red} 164.8} \\ \hline
$R_{\rm n}$(m$\Omega$cm)& {\color{red} 0.80} & {\color{red} 0.57} &{\color{red} 0.42}
&{\color{red} 0.220} &{\color{red} 0.159} & {\color{blue}{\bf 0.097}} \\
\hline \hline
 \end{tabular}
 \label{Table2}
 \end{table}
 
From the above arguments, we expected in the overdoped region a 
different behavior because  $T^*(p)$ diminishes and becomes closer to $T_{\rm c}(p)$
and, concomitantly the SC maximum gap $\Delta_0(p)$ goes down to zero.
In fact, measurements on three overdoped Tl2201 compunds
revealed a decreasing straight line\cite{LscoOver1993} with lower slope than Uemura'line.
A more complete set of data was taken recently on overdoped LSCO compounds
by Bo$\breve{z}$ovi\'c {\it et al}\cite{Bozovic2016} 
and found a similar linear behavior of the old Tl2201 data\cite{LscoOver1993}. 
They developed a technique to grow homogeneous overdoped
films with less  than 1\% variations in $T_{\rm c}$\cite{Bozovic2016,Torchinsky2013}. 
The penetration depth from which $\rho_{sc}(0)$ is derived and 
the resistivity that yields $T_c$ were concomitantly measured, establishing a new 
scaling law: $\rho_{\rm sc}(p,T \sim 0 $ K$)$  is directly proportional to $T_c(p)$. 

To deal with their measurements, we extend our calculations of $\rho_{\rm sc} (p, 0)$ 
to the overdoped LSCO films. Like for the underdoped systems,
${\left < V_{\rm GL}(p,T) \right >}$ has no dimension and we need 
to multiply it by a constant to define the attractive pairing 
potential ${\left <V_{\rm GL}(p,T=0)\right >}$ in eV units. We adjust
${\left < V_{\rm GL}(p_0 = 0.16,T=0)\right >}$ to reproduce the measured 
SC gap  ${\left <\Delta_{\rm sc}(p,0)=0.16)\right >}$,  and all the others 
${\left < V_{\rm GL}(p,0) \right >}$ follow without additional parameters.

\begin{figure}[!ht]
 \includegraphics[height=3.20cm]{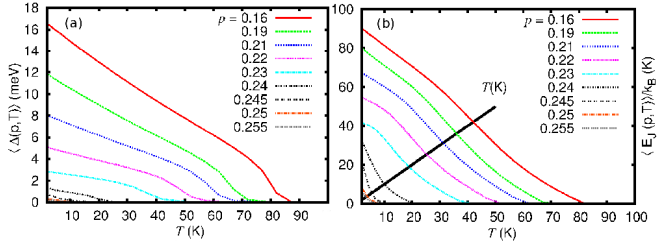}
\caption{ The calculations of $\langle \Delta_{d}(T)\rangle$ and
$\rho_{\rm sc} (T)$ for overdoped LSCO.
(a) ${\left <\Delta_d(p,T)\right >}$ as function of T for 
several compounds. Notice that it remains finite above $T_{\rm c}$. (b) 
The average Josephson energy ${\left < E_{\rm J}(p,T) \right >}/k_{\rm B}$ 
as function of $T$ uses the same resistivity measured in Ref. [\onlinecite{Bozovic2016}]. 
The intersections with $T$ yield $T_{\rm c}(p)$. 
}
\label{fig13ab}
\end{figure}

\begin{table}[!ht] 
\caption{ Properties of overdoped LSCO.
The calculated quantities are shown in red.
$\varepsilon (p)$ is the CH parameter\cite{DeMello2014} that 
accounts for the decreasing PG strength that influences  
$\langle V_{\rm GL}({\bf r})\rangle$ . $V_{\rm GL}(p = 0.16) = 0.247$ eV 
exactly as used in Table I.
$B = 8.0$ from  $R_{\rm n} = B \times \rho_{\rm n}(p = 0.16) = 8.0 \times 0.10)$  
is {\it the only adjustable parameter for the entire overdoped LSCO
series} and with $V_{\rm GL}(p = 0.16)$ are in bold blue. 
The calculate  $\langle \Delta_{d}(0K)\rangle$,
$\rho_{\rm sc} (0)$, $T_{\rm c}$ and the resistivity $\rho_{\rm n}$ are in red. 
The experimental quantities measured in Ref. [\onlinecite{Bozovic2016}] are in parenthesis and in 
black color for comparison. The time of all overdoped simulations were with
$t = 700 \delta t$
}
\centering
\begin{tabular}{|c|c|c|c|c|c|c|}\hline \hline

  $p $ & 0.16  & 0.19 & 0.21  & 0.23 & 0.24  &  0.25    \\ \hline
$\varepsilon (p) $ & {\color{red}0.0133} & {\color{red}0.01364} & {\color{red}0.01379} 
&{\color{red} 0.0139} &{\color{red} 0.01391} & {\color{red}0.01396} \\ \hline
$ -\langle V_{\rm GL}({\bf r})\rangle$ & {\color{red}0.0997} & {\color{red}0.0779} & {\color{red}0.0593 }
& {\color{red}0.0422} & {\color{red}0.0358} & {\color{red}0.0327} \\ \hline
$V_{\rm GL}$ (meV) &{\color{blue} {\bf 234}} & {\color{red}183 }& {\color{red}139} 
& {\color{red}100 } & {\color{red}84 }& {\color{red} 77 } \\ \hline
$\langle \Delta_{d}(0 K)\rangle$ & {\color{red}16.9} & {\color{red} 12.15} & {\color{red} 7.54 }
& {\color{red} 3.19} &{\color{red} 1.47} &{\color{red} 0.33} \\
(meV)\cite{Yoshida2012}  & $(\sim 17)$ &$(\sim 13)$& & & & \\  \hline
$\rho_{\rm sc} (0)$(K) & {\color{red} 90.6} & {\color{red} 80.1} & {\color{red} 66.1} 
& {\color{red} 43.4} & {\color{red} 27.3} & {\color{red} 8.0} \\ \hline
$\rho_{\rm n}$ (m$\Omega$cm)& {\color{blue} {\bf 0.10}} & {\color{red} 0.081} &{\color{red} 0.059}
&{\color{red} 0.038} &{\color{red} 0.028} & {\color{red} 0.016} \\
(Exp.\cite{Bozovic2016}) & (0.10) & (0.08) & (0.065) & (0.036) & (0.022) & (0.01)\\ \hline
$T_{\rm c}$ (K) & {\color{red} 42.2} & {\color{red} 38.0} & {\color{red} 32.3} &  
{\color{red} 21.6} & {\color{red} 13.0} & {\color{red} 4.2} \\
(Estimate\cite{Bozovic2016}) & (41.5) & (37.8) &  (31.1) & (21.2) & (13.0) & (7.9)  \\
\hline \hline

 \end{tabular}
 \label{Table2}
 \end{table}
 
Fig. \ref{fig13ab}(a)  
gives ${\left < \Delta_d(p,T)\right >} \times T$ and for 
$p \lesssim 0.24$ they 
remain finite above $T_{\rm c}$. The low temperature results
are close the experimental $\Delta_0(p)$ values\cite{Yoshida2012} for LSCO 
and are listed in Table II for reference.
Figure \ref{fig13ab}(b) shows ${\left < E_{\rm J}(p,T) \right >} \times T$.   
We perform the same procedure of Ref.[\onlinecite{Mello2019a}] and $R_{\rm n}(p_0 = 0.16)$ is 
adjusted to yield $T_c \sim 42$ K and 
all the others $R_{\rm n}(p)$ follow from their $\rho_{ab}$ experimental ratio. 
The derived $R_{\rm n}(p)$ in this way are listed in Table II and they are 
proportional the measured $\rho_{\rm n}(p)$\cite{Bozovic2016}
$\rho_{\rm n}(p) \sim  \rho_{ab}(p)$ plotted in Ref. [\onlinecite{Bozovic2016}] 
extended data Fig. 8. Again 
from the curves ${\left < E_{\rm J}(p,T) \right >} \times T$ we simultaneously 
derive $T_{\rm c}(p)$ and  $\rho_{\rm sc}(p, 0)$.

It is important to mention that our phase stiffness $\rho_{\rm sc}(p, 0)$ from 
the Josephson coupling is equal their phase stiffness $\rho_{\rm s}(p, 0)$ 
derived from the  magnetic penetration gap at $T = 0$ K. But their $\rho_{\rm s}(p, T)$
is linear with  $T$ and vanishes at $T_{\rm c}(p)$ because there is no Meissner effect
without phase coherence. Our  $\rho_{\rm sc}(p, T)$, the lattice version of the 
superfluid density\cite{Spivak1991}, is equal to  $k_{\rm B}T_{\rm c}(p)$
at $T_{\rm c}(p)$ and  vanishes only when the superconducting fluctuations ($\Delta_d$) 
vanishes, i.e., above $T_{\rm c}(p)$\cite{Ong2010,Gomes2007,Kanigel2008,Dubroka2011,Bilbro2011}.

\begin{figure}[!ht]
\includegraphics[height=5.0cm,angle=0]{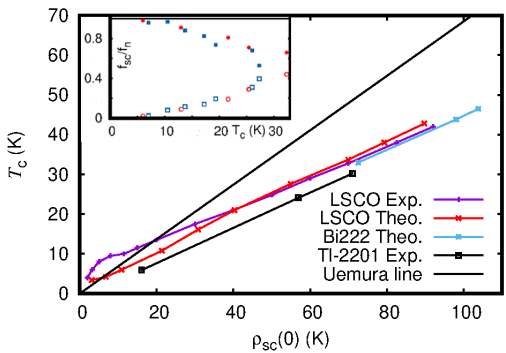}
\caption{ 
The theoretical values of $\rho_{\rm sc}(p,0)$ and $T_{\rm c}(p)$ 
with the experimental results of Ref. [\onlinecite{Bozovic2016}].
We included also three points from Bi2212
($p = 0.16, 0.185$, and 0.22) using the STM data\cite{Gomes2007}
and the Tl2201 results from Ref. [\onlinecite{LscoOver1993}] that, to be plotted together
with LSCO, were divided by 2.2, their ratio of maximum $T_{\rm c}$.  
We draw the Uemura line just for comparison. The inset shows the 
SC fraction $f_{\rm sc}$ (empty red circles) and uncondensed carriers 
$f_{\rm n} = 1 - f_{\rm sc}$ (filled red circles) together with 
the experimental results\cite{Bozovic2019} in empty and 
filled blue squares respectively.
}
\label{fig14}
\end{figure}

The derived $T_c(p) \times \rho_{\rm sc}(p, 0)$ from Fig. \ref{fig13ab}(b) are plotted
in Fig. \ref{fig14} together with the measurements of Bo{\v{z}}ovi{\'c} {\it et al}\cite{Bozovic2016}.
The agreement in the doping range $p = 0.16-0.24$ is almost exact. For $p = 0.24-0.27$
the discrepancy increases with $p$ and it is probably  because the domains of charge
instabilities decrease, the charge distribution becomes almost uniform, 
the SC amplitudes and the Josephson coupling vanish. 

To show that the decreasing of $n_{\rm sc}(p,0)$ in the overdoped regime is universal, 
we included three experimental measurements on Tl2201\cite{LscoOver1993} and three 
calculated Bi2212 points 
($p=0.16, 0.185$, and 0.22) based on STM data\cite{Gomes2007} and 
resistivity measurements\cite{Ando2004} $R_{\rm n}(p)$.
Both Tl2201 and  Bi2212 points are larger
than the LSCO values and were divided by 2.2 (that is their 
$T_{\rm c}(p = 0.16)$ ratio) in order that all the 
${\left < E_{\rm J}(p,T) \right >}$ vs. $T_{\rm c}$ plots be compared in the same
figure. These results indicate that the linear relation 
of overdoped LSCO\cite{Bozovic2016} is common to other cuprates. 
We draw also the Uemura line 
to show that $\rho_{\rm sc}(p,0)$ for overdoped samples are
generally smaller than the underdoped compounds with same $T_{\rm c}(p)$ 
despite the factor of 2 to 3 in the doping level. 

\begin{figure}[!ht]
\includegraphics[height=5.0cm,angle=0]{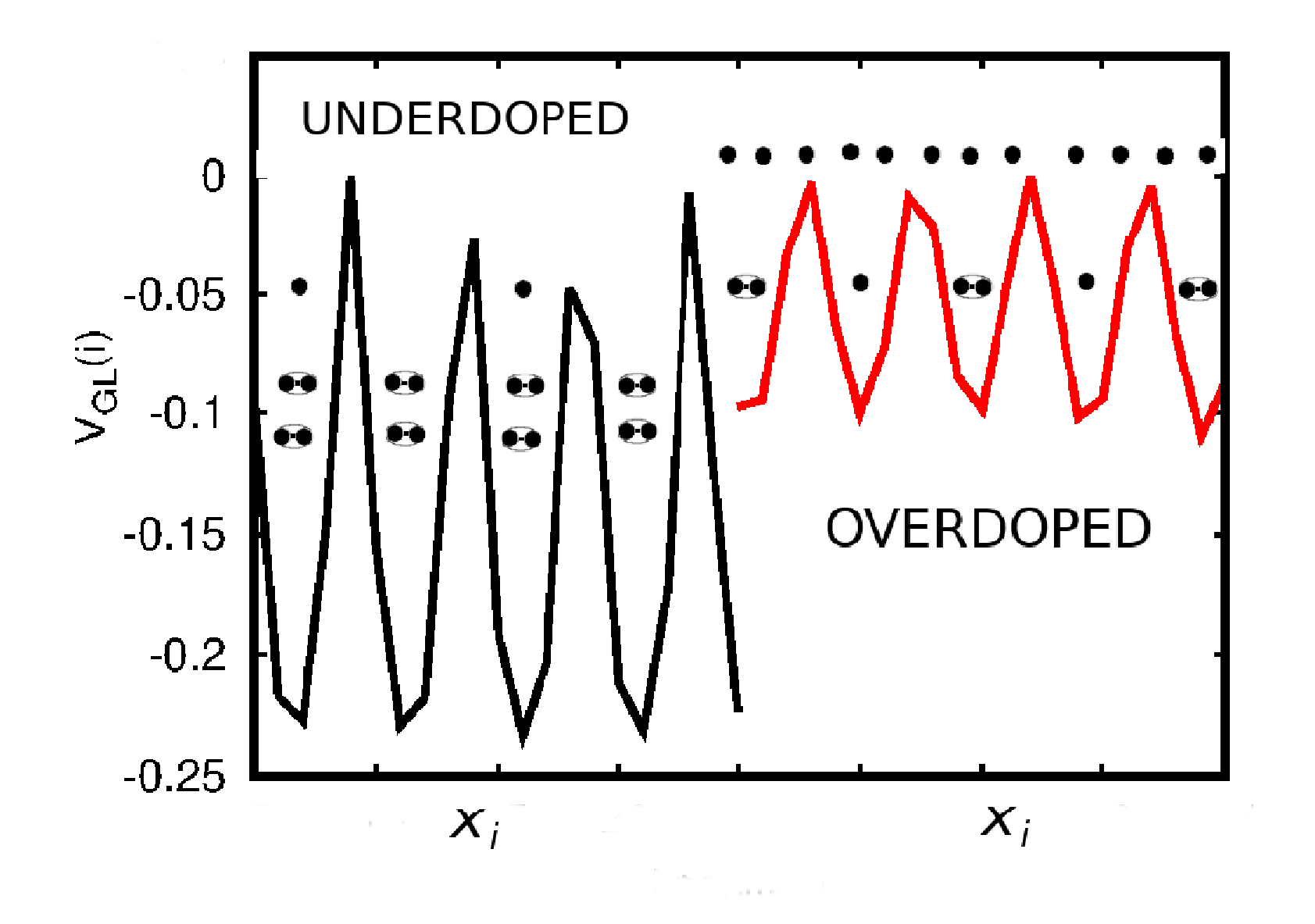}
\caption{ 
Low temperature schematic distribution of the holes and Copper pairs in the presence of the 
$V_{\rm GL}(x_i)$ potential, represented here along the $x$-direction. 
In the underdoped region most holes form SC pairs. In the overdoped region the 
modulation amplitude decreases letting most holes be free carriers and
the superfluid density goes down.
}
\label{fig15}
\end{figure}

Gathering the results from Figs. \ref{figTcxRhoU} and \ref{fig14}, 
and from the Table I and II, we can see that an underdoped compound with
$T_{\rm c} \sim 20$ K has $\rho_{\rm sc} = 25$ K and  hole carriers 
$p = 0.08$, while an overdoped sample with the same
$T_{\rm c} \sim 20$ K has the same $\rho_{\rm sc} = 43$ K 
but with much more carriers $p = 0.23$. 
Thus for a variation of almost 200\% in hole density $p$, 
$\rho_{\rm sc}$ is increased by only 72\% and this
behavior of under/overdoped materials is a common feature. The reason is the
weakening of the SC pair interaction potential $V_{\rm GL}$ with $p$ in
the overdoped as schematically shown in Fig. \ref{fig15}.
Inspecting Figs. \ref{figTcxRhoU} and \ref{fig14} we can see that
the maximum superfluid density occurs at optimal doping where it is likely that
nearly 100\% of the carriers are in the SC state. Taking this as reference,
we can estimate the low temperature superfluid fraction 
of any overdoped compound by the ratio  ${\left < V_{\rm GL}(p)\right >} 
/ {\left < V_{\rm GL}(0.16) \right >}$. 
Some values of $V_{\rm GL}(p)$ are in Table II from which we obtain 
a way to estimate the superfluid fraction 
$f_{\rm sc}(p) = {\left <V_{\rm GL}(p)\right >} /{\left < V_{\rm GL}(0.16)\right >}$ and the uncondensed fraction,
$f_{\rm n}(p) = 1 - f_{\rm sc}(p)$. Plotting $f_{\rm sc}(p)$ vs. $T_{\rm c}$ we
find a parabolic behavior as $T_{\rm c} \rightarrow 0$.

This scenario was confirmed by recent THz optical conductivity 
combined with kHz range mutual inductance measurements\cite{Bozovic2019} 
on overdoped LSCO films.
They observed that the free carriers increases 
as verified by the Drude peak dependence on $p$ at low temperatures while
$\rho_{\rm sc}$ has an opposite behavior\cite{Bozovic2019}. 
In the inset of Fig. \ref{fig14} we plot their spectral weight 
of the superfluid and uncondensed carriers data in empty and 
filled blue squares respectively, normalized by their sum. The calculated SC
fraction $f_{\rm sc}$ (empty red circles) and uncondensed carriers 
$f_{\rm n} = 1 - f_{\rm sc}$ (filled red circles) are also plotted
and the agreement is reasonable.

In their comments\cite{Bozovic2019},
they pointed out that phase separation models could
explain their data if the SC regions would be embedded in a quite large normal volume 
fraction. For the film with $T_{\rm c} = 7$ K, nearly 95\% had
to be in the normal state, 
what was against the uniformity of $T_{\rm c}$ in their films. However, 
our CH-BdG phase separation approach for LSCO produces CDW-like charge modulations 
on the entire system as explained here and before\cite{Mello2019a}. 
The increase of normal carriers is 
due to decreasing amplitude of $V_{\rm GL}(p)$ in 100\% of the system as
shown in the plots for 
$p = 0.016, 0.21$ and 0.25 shown in Fig. 1(c) of 
Ref. [\onlinecite{MelloCM2019}].  The $V_{\rm GL}(p)$ decreases
smoothly with doping diminishes also the values of the SC order parameter amplitude 
$\Delta_d(p)$ (see Table II) and consequently the condensed carrier density, as
schematically shown in Fig. \ref{fig15}.

We conclude pointing out that all the measured  
superfluid density as function of $T_{\rm c}$  are interpreted in a unified way
by the $V_{\rm GL}(p)$ potential that promotes the charge
instabilities and concomitantly, the SC interation. The decreasing of $V_{\rm GL}(p)$,
that is correlated with the PG, reduces the CO constraint and favors free partices
at the same time that the superfluid density is reduced, like  
schematically shown in Fig. \ref{fig15}. 
Together with the quantitative calculations of Fig. \ref{fig14}, we show why the 
superfluid density decreases in the charge
abundant overdoped regime.

I am grateful to I. Bo{\v{z}}ovi{\'c} and J. Tranquada for discussions in the early version
of the manuscript and  acknowledge partial support by the Brazilian agencies CNPq and
FAPERJ.

\end{document}